\documentclass[aps,prc,twocolumn,superscriptaddress,showpacs]{revtex4-2}

\usepackage[dvips]{graphicx} 
\usepackage{color}           
\usepackage{amsmath,amssymb,amsfonts}
\usepackage{bm}
\newcommand{\nc}{\newcommand}           
\nc{\vc}[1]     {\mbox{\boldmath $#1$}} 
\nc{\mapleft}[1]{                       
 \smash{\mathop{                      %
  \hbox to 0.90cm{\rightarrowfill} }\limits_{#1}}}

\nc{\wtil}      {\widetilde}            
\nc{\bra}       {\langle}               
\nc{\ket}       {\rangle}               
\nc{\bras}[1]   {\langle#1|}            
\nc{\kets}[1]   {|#1\rangle}            

\nc{\mydraft}	{\setlength{\topmargin}{-1.5cm}}
\mydraft

\begin{document}

\title{Soft dipole resonance in $^8$C and its isospin symmetry with $^8$He}

\author{Takayuki Myo\footnote{takayuki.myo@oit.ac.jp}}
\affiliation{General Education, Faculty of Engineering, Osaka Institute of Technology, Osaka, Osaka 535-8585, Japan}
\affiliation{Research Center for Nuclear Physics (RCNP), Osaka University, Ibaraki 567-0047, Japan}

\author{Kiyoshi Kat\=o\footnote{kato@nucl.sci.hokudai.ac.jp}}
\affiliation{Nuclear Reaction Data Centre, Faculty of Science, Hokkaido University, Sapporo 060-0810, Japan}

\date{\today}

\begin{abstract}
  We investigate the soft dipole resonance in the proton-rich nucleus $^8_6$C$_2$,
  which is a collective dipole oscillation of four valence protons against the $\alpha$ core,
  and discuss the isospin symmetry with the mirror nucleus $^8_2$He$_6$.
  We use the $\alpha+N+N+N+N$ five-body cluster model and many-body resonances are obtained using the complex-scaling method.
  The $1^-$ resonance of $^8_6$C$_2$ is confirmed at the excitation energy of 13 MeV with a large decay width of 24 MeV, 
  and its structure is similar to the soft dipole resonance in $^8_2$He$_6$, such as the configuration mixing and the spatial properties.
  These results indicate a good isospin symmetry in the soft dipole resonances of two nuclei
  with a common collective excitation of multiproton and multineutron,
  while the ground states of two nuclei show different properties due to the Coulomb repulsion of valence protons in $^8_6$C$_2$, 
  leading to the symmetry breaking. 
  In conclusion, the appearance of the isospin symmetry differs depending on the states of $^8_6$C$_2$ and $^8_2$He$_6$.
\end{abstract}



\maketitle 

\section{Introduction}
Physics of unstable nuclei has been developed with radioactive beam experiments for both neutron-rich and proton-rich sides.
In the neutron-rich side, the neutron halo is known as an exotic structure appearing near the drip-line, such as $^6$He, and $^{11}$Li \cite{tanihata85}.
Many states observed in unstable nuclei can be located above the particle thresholds owing to the weak binding of the excess nucleons.
This property brings the importance of the spectroscopy of resonances to understand the structures of unstable nuclei. 
One of the possible excitations caused by the excess neutrons is the soft dipole resonance,
which is the characteristic excitation of unstable nuclei induced by the collective motion of excess neutrons 
with respect to the stable core nucleus \cite{hansen87,ikeda92,kanungo15}.
From the viewpoint of the isospin symmetry of the nuclear structure, it is an interesting problem to explore the soft dipole resonance
in the proton-rich side as the collective motion of excess protons.

In neutron-rich He isotopes, $^8_2$He$_6$ consists of the $\alpha$ particle and the four excess neutrons with the isospin $T=2$
showing a large neutron-proton ratio of three.
The ground state of $^8$He is known to have a neutron-skin structure of four neutrons with a small separation energy of 3.1 MeV \cite{tilley04}.
Many experiments on $^8$He have been performed to settle the excited states \cite{korsheninnikov93,iwata00,meister02,golovkov09,holl21,lehr22}.
It is interesting to explore the exotic excitations of the excess neutrons and the soft dipole resonance in $^8$He experimentally \cite{lehr22}.

The $^8_6$C$_2$ nucleus is a mirror nucleus of $^8_2$He$_6$ and consists of the $\alpha$ particle and four valence protons with $T=2$.
The $^8$C nucleus is an unbound system with a large proton-neutron ratio of three.
Experimentally, the ground state of $^8$C is observed at 3.4 MeV above the threshold energy of $\alpha+p+p+p+p$ \cite{charity11},
and the excited states have not yet been confirmed.
The comparison of $^8$C and $^8$He as the $T=2$ system is interesting to investigate the effect of the Coulomb interaction in proton-rich nuclei
and get knowledge of the isospin symmetry in unstable nuclei.

Recently we predict the soft dipole resonance in $^8$He and evaluate its effect on the dipole strength function \cite{myo22a,myo22b,myo23}.
We employ the $\alpha+n+n+n+n$ five-body cluster model with the $\alpha$ cluster core and
describe the five-body resonance using the complex scaling method \cite{ho83,moiseyev98,lazauskas05,aoyama06,moiseyev11,myo14a,myo20}
under the correct boundary condition of decaying states.
We investigate the configuration mixing, dipole transition density, and dipole strength distribution for this resonance.
From the results, we interpret the resonance as the soft dipole mode of four valence neutrons ($4n$) oscillating against the $\alpha$ cluster core.
The excitation energy is rather high as 14 MeV, which is understood from the excitation of the relative motion between $\alpha$ and $4n$.

In this paper, we investigate the soft dipole resonance of $^8$C as the mirror state of $^8$He.
We describe many-body unbound states of $^8$C using the $\alpha+p+p+p+p$ five-body cluster model with complex scaling.
We solve the motion of multiproton around the $\alpha$ cluster core in the cluster orbital shell model (COSM) \cite{suzuki88,masui06,masui12,masui14}.
In the COSM, threshold energies of the particle emissions in $^8$C can be reproduced \cite{myo12,myo21},
and this aspect is important to describe the multiproton emissions from $^8$C.
Applying the complex scaling method to the $\alpha+p+p+p+p$ system, we also obtain many-body resonances with complex eigenvalues explicitly.
The effect of resonances can be investigated in the observable, as was done in the case of $^8$He \cite{myo21,myo22a,myo22b}.
We compare the structures of $^8$C with those of $^8$He such as the spatial properties and the Hamiltonian components of resonances,
and discuss the isospin symmetry of the soft dipole resonances in two nuclei.

In Sec.~\ref{sec:method}, we explain the complex-scaled COSM wave function for $^8$C.
In Sec.~\ref{sec:result}, we calculate the $1^-$ states of $^8$C and
investigate the structures of soft dipole resonance of $^8$C in comparison with $^8$He.
A summary is given in Sec.~\ref{sec:summary}.

\section{Method}\label{sec:method}
\subsection{Cluster orbital shell model}

\begin{figure}[b]
\centering
\includegraphics[width=3.5cm,clip]{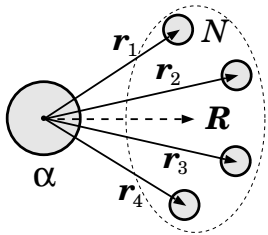}
\caption{Coordinate system of $\alpha+N+N+N+N$ in the COSM for $^8$C and $^8$He.
The vector $\vc{R}$ is a relative coordinate between the center-of-mass of $\alpha$ and the center-of-mass of four valence nucleons.}
\label{fig:COSM}
\end{figure}

We use the five-body cluster orbital shell model (COSM) for $^8$C and $^8$He
assuming the $\alpha$ cluster core \cite{myo22a,myo22b,myo12,myo21}.
The $\alpha$ cluster has a $(0s)^4$ configuration of the harmonic-oscillator basis function
with the length parameter of 1.4 fm to fit the charge radius.
In the COSM, the coordinates of four valence nucleons are measured from the $\alpha$ cluster core and given by $\{\vc{r}_i\}$ with $i=1,\ldots,4$
as is shown in Fig.~\ref{fig:COSM}. 
We use the orthogonality condition model \cite{saito69} to solve the motion of multiproton above the $\alpha$ cluster core.
The total wave function $\Psi^J$ with spin $J$ is expressed by the linear combination of the configurations $\Phi^J_c$ in the COSM as
\begin{eqnarray}
    \Psi^J
&=& \sum_c C^J_c\, \Phi^J_c,
    \qquad
    \Phi^J_c
~=~ \prod_{i=1}^{4} a^\dagger_{\lambda_i}|0\rangle, 
    \label{WF0}
\end{eqnarray}
where the vacuum $|0\rangle$ represents the $\alpha$ cluster core. 
The creation operator $a^\dagger_\lambda$ describes the valence proton with the single-particle state $\lambda$
along the coordinates $\{\vc{r}_i\}$ in Fig.~\ref{fig:COSM}.
The index $c$ represents the set of $(\lambda_1,\lambda_2,\lambda_3,\lambda_4)$ to determine the COSM configuration $\Phi^J_c$.
We consider the available configurations in Eq.~(\ref{WF0}) with a given spin $J$.
In this study, we do not consider the excitation of the $\alpha$ cluster core, which may be necessary to produce the giant dipole resonance. 

The five-body Hamiltonian of $^8$C and $^8$He is the same as used in the previous studies~\cite{myo22a,myo22b,myo12,myo21}:
\begin{eqnarray}
	H
&=&	t_\alpha+ \sum_{i=1}^4 t_i - T_G + \sum_{i=1}^4 v^{\alpha N}_i + \sum_{i<j}^4 v^{NN}_{ij}
    \\
&=&	\sum_{i=1}^4 \left( \frac{\vc{p}^2_i}{2\mu} + v^{\alpha N}_i \right) + \sum_{i<j}^4 \left( \frac{\vc{p}_i\cdot \vc{p}_j}{4 m} + v^{NN}_{ij} \right) .
    \label{eq:Ham}
\end{eqnarray}
We assume the common mass $m$ of a proton and a neutron.
The operators $t_\alpha$, $t_i$, and $T_G$ are the kinetic energies of the $\alpha$ cluster core, one valence nucleon, and the center-of-mass motion, respectively.
The operator $\vc{p}_i$ is the relative momentum between $\alpha$ and a valence nucleon, and $\mu$ is a reduced mass.
The $\alpha$--nucleon interaction $v^{\alpha N}$ is given by the microscopic Kanada-Kaneko-Nagata-Nomoto potential \cite{aoyama06,kanada79},
which consists of the central and $LS$ parts and reproduces the $\alpha$--nucleon scattering data.
The Coulomb part for a valence proton is given by folding the density of the $\alpha$ cluster core \cite{motoba85}.
For nucleon-nucleon interaction $v^{NN}$, we use the Minnesota potential \cite{tang78} for the nuclear part
and the point Coulomb interaction between valence protons.

The single-particle basis function $\phi_\lambda(\vc{r})$ corresponds to the creation operator $a^\dagger_\lambda$
and is expanded with a finite number of Gaussian functions $u_{\ell j}(\vc{r},b)$ in a $jj$ coupling scheme:
\begin{eqnarray}
    \phi_\lambda(\vc{r})
&=& \sum_{q}^{N_{\ell j}} d_{n,q}\; u_{\ell j}(\vc{r},b_{\ell j,q})\, ,
    \label{spo}
    \\
    u_{\ell j}(\vc{r},b_{\ell j,q})
&=& r^{\ell} e^{-(r/b_{\ell j,q})^2/2}\, [Y_{\ell}(\hat{\vc{r}}),\chi^\sigma_{1/2}]_{j}\, ,
    \label{Gauss}
	\\
    \langle \phi_\lambda | \phi_{\lambda'} \rangle 
&=& \delta_{\lambda,\lambda'}
~=~ \delta_{n,n'}\, \delta_{\ell,\ell'}\, \delta_{j,j'}.
    \label{Gauss2}
\end{eqnarray}
The index $q$ is to set the length parameter $b_{\ell j,q}$ with a number of $N_{\ell j}$,
and $n$ is the index to distinguish the different radial functions, and then $\lambda$ is a set of $(n,\ell,j)$.
The coefficients $\{d_{n,q}\}$ are determined from the orthonormalization of the basis states $\{\phi_\lambda\}$ in Eq.~(\ref{Gauss2}).
We use $N_{\ell j}=12$ at most and the range of $b_{\ell j,q}$ is taken from 0.3 fm to around 40 fm.
The multinucleon configuration $\Phi^J_c$ in Eq.~(\ref{WF0}) is given by the products of $\phi_\lambda(\vc{r})$.

For $\phi_\lambda(\vc{r})$, we consider the orbital angular momenta $\ell\le 2$ for the positive-parity states of $^8$C and $^8$He \cite{myo21}.
In $v^{NN}$, we reduce the repulsive strength of the Minnesota potential from 200 MeV to 173.7 MeV
to fit the two-neutron separation energy of $^6$He as 0.98 MeV \cite{myo12,myo21}.
In this Hamiltonian, the ground-state energy of $^8$He is obtained as $-3.21$ MeV measured from the threshold energy of $\alpha+n+n+n+n$, 
which is close to the experimental value of $-3.11$ MeV.
For $1^-$ states of $^8$He, we calculated the electric dipole transition from the ground state
by adding the configurations with the $\ell=3$ states \cite{myo22a,myo22b}.
This treatment nicely describes the sum-rule value of the dipole transition to check the completeness relation of the $1^-$ states. 
In this study, we follow the same condition to calculate the $1^-$ states of $^8$C.

\subsection{Complex scaling method}
We describe many-body resonance with the complex scaling \cite{ho83,moiseyev98,moiseyev11,aoyama06,myo14a,myo20}.
In the complex scaling, the relative coordinates and momenta $\{\vc{r}_i,\vc{p}_i\}$ are transformed using a common and real scaling angle $\theta$:
\begin{eqnarray}
\vc{r}_i \to \vc{r}_i\, e^{i\theta},\qquad
\vc{p}_i \to \vc{p}_i\, e^{-i\theta}.
\end{eqnarray}
The complex-scaled Schr\"odinger equation is given with the complex-scaled Hamiltonian $H_\theta$ as 
\begin{eqnarray}
	H_\theta   \Psi^J_\theta
&=&     E^J_\theta \Psi^J_\theta ,
	\label{eq:eigen2}
        \\
    \Psi^J_\theta
&=& \sum_c C^J_{c,\theta}\, \Phi^J_c .
    \label{eq:WF_CSM}
\end{eqnarray}
We calculate the complex-scaled Hamiltonian matrix elements using the COSM basis states $\Phi^J_c$
and solve the eigenvalue problem of Eq.~(\ref{eq:eigen2}).
We obtain the energy eigenvalues $E^J_\theta$ and the coefficients $\{C^J_{c,\theta}\}$,
which determine the complex-scaled wave function $\Psi^J_\theta$ in Eq.~(\ref{eq:WF_CSM}),
where $\sum_c \{C^J_{c,\theta}\}^2 =1$ for the normalization of the eigenstates.

In this study, the energy eigenvalues $E^J_\theta$ are discretized for bound, resonant,and continuum states,
measured from the threshold energy of $\alpha+N+N+N+N$,
and obtained on a complex energy plane according to the so-called ABC theorem \cite{ABC}.
In the complex scaling, the Riemann branch cuts are rotated down by $2\theta$ in the complex energy plane,
and start from the corresponding threshold energies of the cluster emissions.
The continuum states are obtained on these rotated $2\theta$ lines.
The bound and resonant states are physical solutions and their eigenenergies are independent of $\theta$.
The Hermitian product is not used in the biorthogonal relation \cite{berggren68},
which is conventionally treated as the C-product \cite{moiseyev98} in the complex-scaled matrix elements;
one does not take the complex conjugate in the radial component of the bra state.
In the complex scaling, one can identify the resonances in the complex energy plane
using the stationary property of the resonance eigenenergies 
in the trajectory with respect to the scaling angle $\theta$ \cite{ho83,moiseyev98,aoyama06}.

\section{Results}\label{sec:result}

\subsection{Energies}
\begin{figure}[t]
\centering
\includegraphics[width=8.0cm,clip]{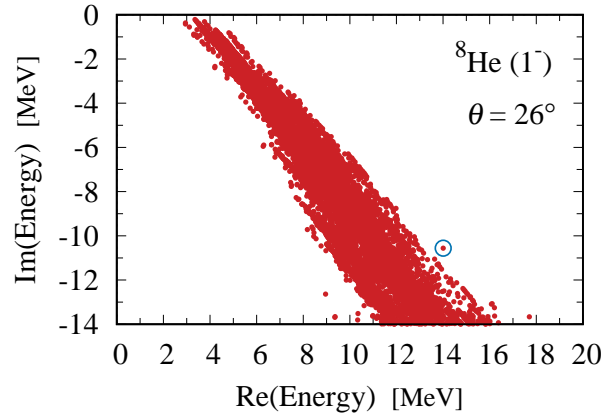}
\caption{
  $1^-$ eigenstates of $^8$He using the complex scaling with $\theta=26^\circ$ in the complex energy plane measured from the ground-state energy.
  The blue circle indicates the eigenenergy of the soft dipole resonance (SDR). The results are taken from Ref. \cite{myo22b}.
}
\label{fig:ene_8He}
\end{figure}
\begin{figure}[t]
\centering
\includegraphics[width=8.0cm,clip]{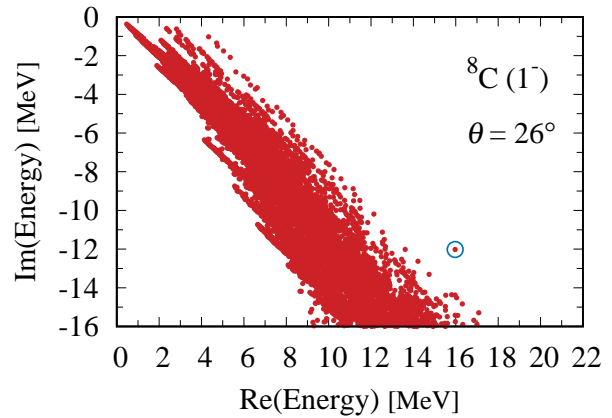}
\caption{
  $1^-$ eigenstates of $^8$C using the complex scaling with $\theta=26^\circ$ in the complex energy plane measured from the $\alpha+p+p+p+p$
  threshold energy.
  The blue circle indicates the eigenenergy of the soft dipole resonance (SDR).
}
\label{fig:ene_8C}
\end{figure}

We discuss the soft dipole resonances of $^8$C and $^8$He, where some of the results of $^8$He are already reported in Refs. \cite{myo22a,myo22b}.
For reference, in Fig. \ref{fig:ene_8He}, we show the energy eigenvalues of $^8$He ($1^-$) using the complex scaling
in the complex energy plane, measured from the ground state \cite{myo22b}.
We use the scaling angle $\theta=26^\circ$, which gives the stationary point of the $1^-$ resonance energy in the complex energy plane.
The discretized continuum states are obtained almost along the $2\theta$ lines, starting from the energy eigenvalues of the subsystems of $^8$He.
We confirm the resonance pole at $(E_x,\Gamma)=(14.02,~21.12)$ MeV and
that this resonance exhausts about half of the electric dipole transition from the ground state \cite{myo22a,myo22b}.
We have also investigated the dipole transition density of this resonance, and 
the results suggest the interpretation of the dipole oscillation of four valence neutrons against the $\alpha$ core.
Hence we regard this resonance as the soft dipole resonance (SDR).

For the configurations of the SDR in $^8$He, we list the dominant part of the squared amplitudes $(C_{c,\theta}^J)^2$
being the complex numbers in Table \ref{tab:config}.
In each configuration, we sum the values of the different radial components of valence neutrons in the same orbit of $(\ell, j)$.
The results show a mixing of the various configurations with the excitations of multineutron from the $(p_{3/2})^4$ configuration,
that is dominant in the ground state of $^8$He \cite{myo21}.
The property of the strong mixing agrees with the collective excitation of four neutrons in the SDR.

\begin{figure}[t]
\centering
\includegraphics[width=5.0cm,clip]{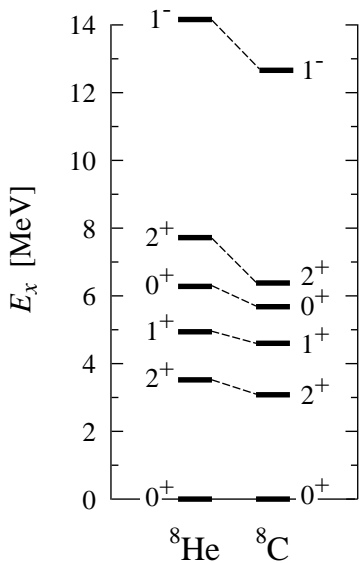}
\caption{Comparison of the excitation energy spectra between $^8$He an $^8$C in units of MeV.}
\label{fig:mirror}
\end{figure}

Next, we discuss the results of $^8$C.
The ground state of $^8$C is a $0^+$ resonance obtained at 3.32 MeV measured from the $\alpha+p+p+p+p$ threshold energy \cite{myo12,myo21},
which is close to the experimental value of 3.449(30) \cite{charity11}, and the decay width is obtained as 0.072 MeV. 
Similarly to the SDR in $^8$He, we investigate the $1^-$ states of $^8$C using the complex scaling.
In Fig. \ref{fig:ene_8C}, we show the energy eigenvalues of $^8$C measured from the threshold energy of $\alpha+p+p+p+p$
using the scaling angle $\theta=26^\circ$, which is the same value as used for $^8$He.
As well as the discretized continuum states, we confirm the resonance pole at $(E_r, \Gamma)=(15.97,~24.04)$ MeV with $E_x=12.65$ MeV, which is a mirror state of the SDR in $^8$He.
The excitation energy is lower than that of $^8$He by 1.4 MeV and the decay width is larger than that of $^8$He by 2.9 MeV.
The dominant configurations of the $1^-$ resonance of $^8$C are shown in Table \ref{tab:config} as well as those of $^8$He.
A similar tendency is confirmed in the mixing property of the configurations between the resonances in two nuclei indicating the isospin symmetry.

\begin{table}[t]
  \caption{Dominant configurations with the squared amplitudes $(C_{c,\theta}^J)^2$ of the soft dipole resonances in $^8$He and $^8$C.}
  \label{tab:config}
  \centering
\begin{ruledtabular}
\begin{tabular}{c|cc}
                                         &  $^8$He              &  $^8$C            \\ \cline{2-3}
Configuration                            &  $(C_{c,\theta}^J)^2$&  $(C_{c,\theta}^J)^2$  \\ \hline
$(p_{3/2})^3(d_{5/2})$                   &  $0.573-0.186i$      &  $0.666-0.165i$   \\  
$(p_{3/2})^2(p_{1/2})(d_{3/2})$          &  $0.197+0.118i$      &  $0.138+0.115i$   \\  
$(p_{3/2})^3(d_{3/2})$                   &  $0.149+0.120i$      &  $0.087+0.150i$   \\  
$(p_{3/2})^2(d_{5/2})(f_{7/2})$          &  $0.027+0.007i$      &  $0.025+0.002i$   \\  
$(p_{3/2})^2(p_{1/2})(d_{5/2})$          &  $0.022-0.012i$      &  $0.033-0.020i$   \\  
$(p_{3/2})(p_{1/2})^2(d_{3/2})$          &  $0.018-0.064i$      &  $0.018-0.078i$   \\  
$(p_{3/2})(d_{5/2})^3$                   &  $0.015-0.003i$      &  $0.016-0.005i$   \\  
$(p_{3/2})(d_{3/2})^2(d_{5/2})$          &  $0.007-0.001i$      &  $0.008-0.002i$   \\  
\end{tabular}
\end{ruledtabular}
\end{table}

In Fig. \ref{fig:mirror}, we compare the excitation energy spectra of $^8$He and $^8$C using their resonance energies, 
where we include the positive-parity states reported in Ref. \cite{myo21}.
The order of the energy levels are same, but the level spacing is smaller in $^8$C than that in $^8$He including SDRs.
This is an indication of the isospin-symmetry breaking from the Thomas-Ehrman effect of the Coulomb interaction. 
The level spacing is related to the larger decay widths of the resonances in $^8$C than those of $^8$He
as shown in Figs, \ref{fig:ene_8He} and \ref{fig:ene_8C}, and as discussed in Ref. \cite{myo21}.
When we calculate the distance between the complex energy eigenvalues from the $0^+$ ground state to the SDR in the complex energy plane,
$^8$He and $^8$C give 17.6 MeV and 17.4 MeV, respectively, which are close to each other.

We discuss the excitation energies $E_x$ of SDRs in $^8$He and $^8$C because they are more than 10 MeV and not small.
We interpret these resonances as the collective dipole oscillation of four valence nucleons ($4N$) against the $\alpha$ core.
Hence it is meaningful to consider the relative motion between the $\alpha$ core and $4N$.
In the ground states of two nuclei, the relative distances $(\bra \vc{R}^2 \ket)^{1/2}$ between the $\alpha$ core and $4N$ are
2.05 fm for $^8$He and 2.36 fm for $^8$C, respectively, where the vector $\vc{R}$ is defined in Fig. \ref{fig:COSM}.
When we assume the lowest $(0p)^4$ configuration of $4N$ in $^8$He and $^8$C using the harmonic-oscillator basis states,
this configuration leads to the quanta $N_R=2$ in the $\alpha$-$4N$ relative motion, 
while $4N$ has the lowest internal quanta of two due to the Pauli principle.
We can estimate $\hbar\omega_R$ of the relative motion using the length parameter $b_R$ and the reduced mass $\mu_R$ as follows
\begin{eqnarray}
  \hbar\omega_R &=& \frac{\hbar^2}{\mu_R\,b_R^2},
  \qquad
  \langle \vc{R}^2 \rangle
  ~=~  \left( \frac{3}{2}+N_R \right) b_R^2.
\end{eqnarray}
The value of $\hbar\omega_R$ becomes the reference for the excitation energy of the relative motion.
For $^8$He, this relation gives $\hbar\omega_R=17.2$ MeV while the calculation gives 14.0 MeV.
For $^8$C, $\hbar\omega_R=13.0$ MeV and the calculation gives 12.7 MeV.
From these comparisons, the estimated values of $\hbar\omega_R$ can simulate well the excitation energies of the resonances obtained in two nuclei.
These results support the picture of the excitation of the relative motion in SDRs of two nuclei, namely the collective excitations of $4N$.
The obtained excitation energies are also slightly lower than the estimated ones,
which indicates a deviation from the simple $\alpha+4N$ two-body model.

\subsection{Radii}

\begin{table}[t]  
  \caption{
    Various radii of the ground $0^+$ states and the soft dipole resonances (SDR) of $^8$He and $^8$C
    for matter, proton, and neutron parts in units of fm.
    Mean distances between the $\alpha$ core and one valence nucleon $\alpha$-$N$, and
    between the $\alpha$ core and four valence nucleons $\alpha$-$4N$,
    and the radii of four valence nucleons $4N$ are also shown in fm.}
  \label{tab:radius}
  \centering
  \begin{ruledtabular}
    \begin{tabular}{c|cc|cc}
                  & \multicolumn{2}{c}{$^8$He} & \multicolumn{2}{c}{$^8$C}      \\ \cline{2-5}
                  & ground & SDR               & ground        & SDR            \\ \hline
      matter      & 2.53   & $3.11+0.86i$      & $2.81 -0.08i$ & $3.09 + 1.27i$ \\  
      proton      & 1.81   & $1.97+0.29i$      & $3.06 -0.10i$ & $3.39 + 1.46i$ \\
      neutron     & 2.73   & $3.41+0.99i$      & $1.90 -0.01i$ & $1.94 + 0.39i$ \\
$\alpha$-$N$      & 3.56   & $4.58+1.42i$      & $4.05 -0.12i$ & $4.54 + 2.05i$ \\ 
$\alpha$-$4N$     & 2.05   & $2.67+0.84i$      & $2.36 -0.03i$ & $2.62 + 1.15i$ \\
$4N$              & 2.91   & $3.72+1.14i$      & $3.29 -0.13i$ & $3.71 + 1.70i$ \\ 
\end{tabular}
\end{ruledtabular}
\end{table}

For the comparison of $^8$He and $^8$C, we discuss the spatial properties of the SDRs in two nuclei as well as the ground states.
In Table~\ref{tab:radius}, we list the various radii and relative distances for $^8$He and $^8$C.
The value of $4N$ is the root-mean-square radius of $4N$ measured from the center of mass of $4N$.
The value of $\alpha$-$4N$ is the relative distance between the center of mass of the $\alpha$ core and the center of mass of $4N$.
Note that, for resonances, the radius becomes complex numbers in general \cite{myo20,myo14b},
and we recently propose a general scheme to interpret the complex expectation values for resonances \cite{myo23},
in which the real part can be related to the physical interpretation.
According to this scheme, we discuss the real part of the complex radii in the present analysis.

In the ground states, it is found that the matter radius of $^8$C is larger than that of $^8$He by about 0.28 fm.
This difference comes from the large proton radius (3.06 fm) of $^8$C than the neutron radius (2.73 fm) of $^8$He.
This is due to the Coulomb repulsion in protons, which breaks the symmetry between two nuclei.
In addition, the $^8$C ground state is a resonance located above the $\alpha+p+p+p+p$ threshold energy.
The $\alpha$-$N$ distance in $^8$C (4.05 fm) is also larger than that of $^8$He (3.56 fm).
The radius of four protons ($4p$) is larger than that of four neutrons ($4n$) by about 0.38 fm.
Hence the ground state of $^8$C is spatially expanded in comparison with $^8$He due to the Coulomb repulsion,
and in this aspect the isospin symmetry is broken in the ground states of two nuclei.
Note that the amplitudes of the dominant configurations are similar between the ground states of $^8$He and $^8$C,
indicating the isospin symmetry \cite{myo21}, although their spatial sizes are different.

For the SDRs in two nuclei, it is found that two resonances show very similar values in Table \ref{tab:radius},
which is quite different from the results of the ground states.
In every component, we can confirm a good correspondence such as between the proton radius in $^8$C and the neutron radius of $^8$He,
and between the radii of $4p$ and $4n$.
These results indicate that the isospin symmetry is restored in the SDRs of two nuclei.
This can come from the common spatial structure of two resonances as the collective dipole oscillation of $4N$ against the $\alpha$ core.
Note that the imaginary parts are larger in $^8$C than those of $^8$He for every component.

In the summary of spatial structures of two nuclei, the isospin symmetry is broken in the ground states due to the Coulomb repulsion.
On the other hand, the symmetry is restored in the SDRs, which indicates the different role of the Coulomb interaction in the resonances.
So far, we have shown the cases where the radius of proton-rich nuclei can be smaller than the mirror neutron-rich nuclei 
for the excited resonances \cite{myo23,myo14b}.
This is understood from the barrier effect of Coulomb interaction in the excited states.

\subsection{Hamiltonian components}

\begin{table}[t]  
  \caption{Hamiltonian components of the ground $0^+$ states of $^8$He and $^8$C in units of MeV.
    The values are measured from the threshold energy of $\alpha+N+N+N+N$.}
  \label{tab:ham1}
  \centering
  \begin{ruledtabular}
    \begin{tabular}{l | r rr}
   & \multicolumn{1}{c}{$^8$He ($0^+_1$)}  & \multicolumn{1}{c}{$^8$C ($0^+_1$)}  \\ \hline
     Total energy     & $- 3.21$   & $~~3.32 - 0.036i$  \\
     Kinetic energy   & $ 60.31$   & $ 48.90 - 0.77 i$  \\
$\alpha$-$N$: central & $-46.87$   & $-38.51 + 0.55 i$  \\
$\alpha$-$N$: $LS$    & $- 6.97$   & $- 5.64 + 0.10 i$  \\
$\alpha$-$N$: Coulomb &  --        & $  3.76 - 0.04 i$  \\
   $NN$: nuclear      & $- 9.68$   & $- 7.44 + 0.16 i$  \\  
   $NN$: Coulomb      &  --        & $  2.25 - 0.03 i$  \\
\end{tabular}
\end{ruledtabular}
\end{table}

\begin{table}[b]  
  \caption{Hamiltonian components of soft dipole resonance in $^8$He in units of MeV.
  The values are measured from the threshold energy of $\alpha+n+n+n+n$.
  The absolute values are also shown.}
  \label{tab:ham2}
  \centering
  \begin{ruledtabular}
    \begin{tabular}{l|rrr}
                       & $^8$He ($1^-$)~~~  & absolute\\ \hline
     Total energy      & $ 10.81- 10.56i$   & 15.11   \\
     Kinetic energy    & $ 43.45- 24.26i$   & 49.77   \\
$\alpha$-$N$: central  & $-21.24+\,~7.55i$  & 22.55   \\
$\alpha$-$N$: $LS$     & $- 3.19+\,~0.15i$  &  3.19   \\
   $NN$: nuclear       & $- 8.22+\,~6.00i$  & 10.17   \\
\end{tabular}
\end{ruledtabular}
\end{table}

According to the results of the radii in $^8$C and $^8$He,
finally, we discuss the Hamiltonian components of the ground states and SDRs in two nuclei,
which are measured from the threshold energy of $\alpha+N+N+N+N$. 
In the interaction components, we take a summation over four valence nucleons,
and the $\alpha$-$N$ interaction is decomposed into the central, $LS$, and Coulomb parts,
and the $NN$ interaction is decomposed into the nuclear and Coulomb parts.

In Table~\ref{tab:ham1}, we list the Hamiltonian components of the ground states of two nuclei.
For $^8$C, the ground state is a resonance and the Hamiltonian components become complex numbers.
We compare the real parts of them with those of $^8$He because the imaginary parts are relatively smaller than the real parts.
The total energy of $^8$C subtracting the Coulomb components is $-2.7$ MeV, which is repulsive in comparison with $-3.2$ MeV of $^8$He by 0.5 MeV. 
In the comparison of two nuclei, there are large differences in each of the Hamiltonian components;
the kinetic energy is 11.4 MeV, and the $\alpha$-$N$ central interaction energy is 8.4 MeV. 
As well as the radii shown in Table \ref{tab:radius}, 
these results represent the differences in spatial spread in the ground states of $^8$He and $^8$C.
In the ground states, $^8$He is a bound state and $^8$C is a resonance due to the Coulomb repulsion,
and then the valence protons in $^8$C become distributed more widely than the valence neutrons in $^8$He. 
This spatial effect reduces the kinetic energy of $^8$C from that of $^8$He.
The magnitude of the interaction energy is also reduced in $^8$C
because of the increase of the amplitudes of the valence protons outside the interaction region.

\begin{table}[t]  
  \caption{Hamiltonian components of soft dipole resonance in $^8$C in units of MeV.
  The values are measured from the threshold energy of $\alpha+p+p+p+p$.
  The absolute values are also shown.}
  \label{tab:ham3}
  \centering
  \begin{ruledtabular}
    \begin{tabular}{l|rrr}
                       & $^8$C ($1^-$)~~~     & absolute\\ \hline
     Total energy      & $ 15.97 - 12.02i$  & 19.99   \\
     Kinetic energy    & $ 38.22 - 27.91i$  & 47.32   \\
$\alpha$-$N$: central  & $-17.96 + 10.03i$  & 20.57   \\
$\alpha$-$N$: $LS$     & $- 2.83 +\,~0.54i$ &  2.88   \\
$\alpha$-$N$: Coulomb  & $  3.20 -\,~0.91i$ &  3.32   \\
   $NN$: nuclear       & $- 6.51 +\,~6.97i$ &  9.53   \\
   $NN$: Coulomb       & $  1.86 -\,~0.75i$ &  2.01   \\
\end{tabular}
\end{ruledtabular}
\end{table}

In Tables~\ref{tab:ham2} and \ref{tab:ham3}, we list the Hamiltonian components of the SDRs in $^8$He and $^8$C, respectively.
They are complex values with non-negligible imaginary parts, which are larger in $^8$C than $^8$He,
and then we list their absolute values for reference.
The total energy of $^8$C subtracting the Coulomb components is $10.91-10.36i$ MeV, which is very close to $10.81- 10.56i$ MeV of $^8$He.
In the absolute values, it is found that the individual components are close to each other in two nuclei.
The difference in the kinetic energies is 2.5 MeV, which is much smaller than 11.4 MeV in the ground states.
The difference in the $\alpha$-$N$ central interaction energy is 2.0 MeV, which is smaller than 8.4 MeV in the ground states.
The difference in the $NN$ central interaction energy is 0.6 MeV, which is also smaller than 2.3 MeV in the ground states.
From these comparisons, it is found that two SDRs show a similar structure and keep a good isospin symmetry.
This is the same conclusion as confirmed for radii of SDRs shown in Table \ref{tab:radius}.

In the overall analysis via the spatial properties and Hamiltonian components in two nuclei,
the SDRs in $^8$He and $^8$C retain the isospin symmetry, which is different from the ground states.
This result indicates that the SDRs in two nuclei are in a common excitation mode of the collective dipole oscillation of $4N$ against the $\alpha$ cluster core.
In the future, it would be interesting to examine this result allowing the excitation of the $\alpha$ cluster,
which can bring the coupling of the SDR with the giant dipole resonance.

\section{Summary}\label{sec:summary}
Recently, we have predicted the soft dipole resonance in $^8$He contributing largely to the electric dipole strength \cite{myo22a,myo22b}.
This state is regarded as the collective dipole oscillation of four valence neutrons against the $\alpha$ cluster core.
According to this result, in this paper, we investigated the soft dipole resonance in the mirror nucleus $^8$C 
and discussed the isospin symmetry in the mirror resonances of $^8$He and $^8$C.

The $^8$C nucleus is the unbound system and
we use the complex scaling method to obtain the five-body resonances of $^8$C in the $\alpha+p+p+p+p$ cluster model.
We successfully obtained the dipole resonance of $^8$C with the excitation energy of 13 MeV and a large decay width of 24 MeV.
This resonance has a similar property of the configuration mixing to that of $^8$He and then we assign this state as the soft dipole resonance.
This resonance can correspond to the dipole oscillation of four valence protons against the $\alpha$ cluster core.
The excitation energy is reasonably explained in a picture of the excitation of the relative motion between four protons and the $\alpha$ cluster core.

We compare the soft dipole resonances in $^8$C and $^8$He with several quantities.
The spatial properties such as radii show similar values between two resonances,
while the ground states do not show the symmetry in their radii.
We also compare the Hamiltonian components in two nuclei and the good symmetry is confirmed in the soft dipole resonances.
In conclusion, the isospin symmetry is broken in the ground states of two nuclei, which comes from the Coulomb repulsion.
On the other hand, the isospin symmetry is restored in the soft dipole resonances of two nuclei.
This is because two resonances are in a common excitation mode of the collective dipole oscillation of four valence nucleons against the $\alpha$ cluster core.
The present analyses show that the appearance of the isospin symmetry differs depending on the states of $^8$C and $^8$He.

The present theoretical prediction of the soft dipole resonances in $^8$C and $^8$He would be examined in future experiments \cite{lehr22}.
It is also interesting to explore other candidates of drip-line nuclei to investigate the isospin symmetry in collective excitations.

\section{Acknowledgments}
This work was supported by JSPS KAKENHI Grants No. JP18K03660, No. JP20K03962, and No. JP22K03643.
Numerical calculations were partly achieved through the use of SQUID at the Cybermedia Center, Osaka University.


\section*{References}
\def\JL#1#2#3#4{ {{\rm #1}} \textbf{#2}, #4 (#3)}  
\nc{\PR}[3]     {\JL{Phys. Rev.}{#1}{#2}{#3}}
\nc{\PRC}[3]    {\JL{Phys. Rev.~C}{#1}{#2}{#3}}
\nc{\PRA}[3]    {\JL{Phys. Rev.~A}{#1}{#2}{#3}}
\nc{\PRL}[3]    {\JL{Phys. Rev. Lett.}{#1}{#2}{#3}}
\nc{\NP}[3]     {\JL{Nucl. Phys.}{#1}{#2}{#3}}
\nc{\NPA}[3]    {\JL{Nucl. Phys.}{A#1}{#2}{#3}}
\nc{\PL}[3]     {\JL{Phys. Lett.}{#1}{#2}{#3}}
\nc{\PLB}[3]    {\JL{Phys. Lett.~B}{#1}{#2}{#3}}
\nc{\PTP}[3]    {\JL{Prog. Theor. Phys.}{#1}{#2}{#3}}
\nc{\PTPS}[3]   {\JL{Prog. Theor. Phys. Suppl.}{#1}{#2}{#3}}
\nc{\PRep}[3]   {\JL{Phys. Rep.}{#1}{#2}{#3}}
\nc{\JP}[3]     {\JL{J. of Phys.}{#1}{#2}{#3}}
\nc{\PPNP}[3]   {\JL{Prog. Part. Nucl. Phys.}{#1}{#2}{#3}}
\nc{\PTEP}[3]   {\JL{Prog. Theor. Exp. Phys.}{#1}{#2}{#3}}
\nc{\andvol}[3] {{\it ibid.}\JL{}{#1}{#2}{#3}}

\end{document}